# Data Security and Privacy Protection in Public Cloud

Yue Shi
Viterbi School of Engineering
University of Southern California
Los Angeles, CA
yueshi@usc.edu

*Abstract*—This paper discusses about the challenges, advantages and shortcomings of existing solutions in data security and privacy in public cloud computing. As in cloud computing, oceans of data will be stored. Data stored in public cloud would face both outside attacks and inside attacks since public cloud provider themselves are untrusted. Conventional encryption could be used for storage, however most data in cloud needs further computation. Decryption before computation will cause large overheads for data operation and lots of inconvenience. Thus, efficient methods to protect data security as well as privacy for large amount of data in cloud are necessary.

In the paper, different mechanisms to protect data security and privacy in public cloud are discussed. A data security and privacy enabled multi-cloud architecture is proposed.

*Keywords— Cloud computing, data confidentiality, data privacy, data integrity.*

I. INTRODUCTION

Cloud computing has become an emerging technique due to its on demand service and scalablity features. Most usage of cloud today is in data storage and big data or computation-intensive applications. Thus data security and privacy has become the chief concern, especially for business level data. Data security mainly includes data confidentiality, availa-bility and integrity. Data privacy is to prevent identification of data stored in cloud. According to [3], data security and privacy issues in cloud exist during the data life cycle from generation, transfer, use, share, storage, archival until destruction.

Traditional methods for data security usually rely on data encryption and access control. Data encryption with AES or other encryption methods would prevent valuable infor-mation leakage although the adversary gets hold of the data. However it has efficiency issue when dealing with oceans of data in cloud environment due to large encryption and de-cryption overhead in storage and computation.

Access control is to prevent unauthorized users to access data. However, in cloud computing, users do not have physi-cal control over the machines they store data on, and also the same physical machine could be shared by multiple tenants with virtualization, adversary would be able to monitor the physical machine behavior to obtain valuable data from other tenants [10], and also the cloud providers themselves are unreliable, they might accidentally or intentionally mod-ify or leak the data stored to adversaries.

In terms of the particularities of data security in cloud computing enviroment, many researches have been done. This paper focuses on the protection of the data confidentiali-ty in different phases of data life cycle. The paper is organized as follows.

In Section II, the cloud architecture and possible attacks in different points are discussed. In Section III, data security con-cern and possible attacks during various stages of data life cycle in cloud environment will be discussed. In section IV, V and VI, various protection methods of data confidentiality, availability, integrity and privacy against different kinds of attacks will be discussed in details, problems and ad-vantages of these different techniques are compared. A new cloud data security and privacy enabled architecture and confidentiality ranking system are propsed in section VII..

.

II. CLOUD ARCHITECTURE AND SECURITY ISSUES

In public cloud environment, threats come from both the outsider and insider attack. Fig.1 shows the cloud architecture and attacks. The outsider attacks by malicious codes, DDoS attack, network eavesdropping etc. There are three layers in cloud computing platform. In the infrastructure layer, each physical machine has multiple virtual machines (VMs) installed. The platform layer provides the platform for customers. Customers could have their own applications or softwares and configurations installed. And the software layer provides the software stacks by the cloud providers.

For the client side, a customer could either be a legal user or an attacker pretending as legal users. Network eavesdroppers could also sit in between to perform man in the middle attacks. Firewalls or Intrusion Detection Systems (IDS) could be installed to protect the entire cloud environment.

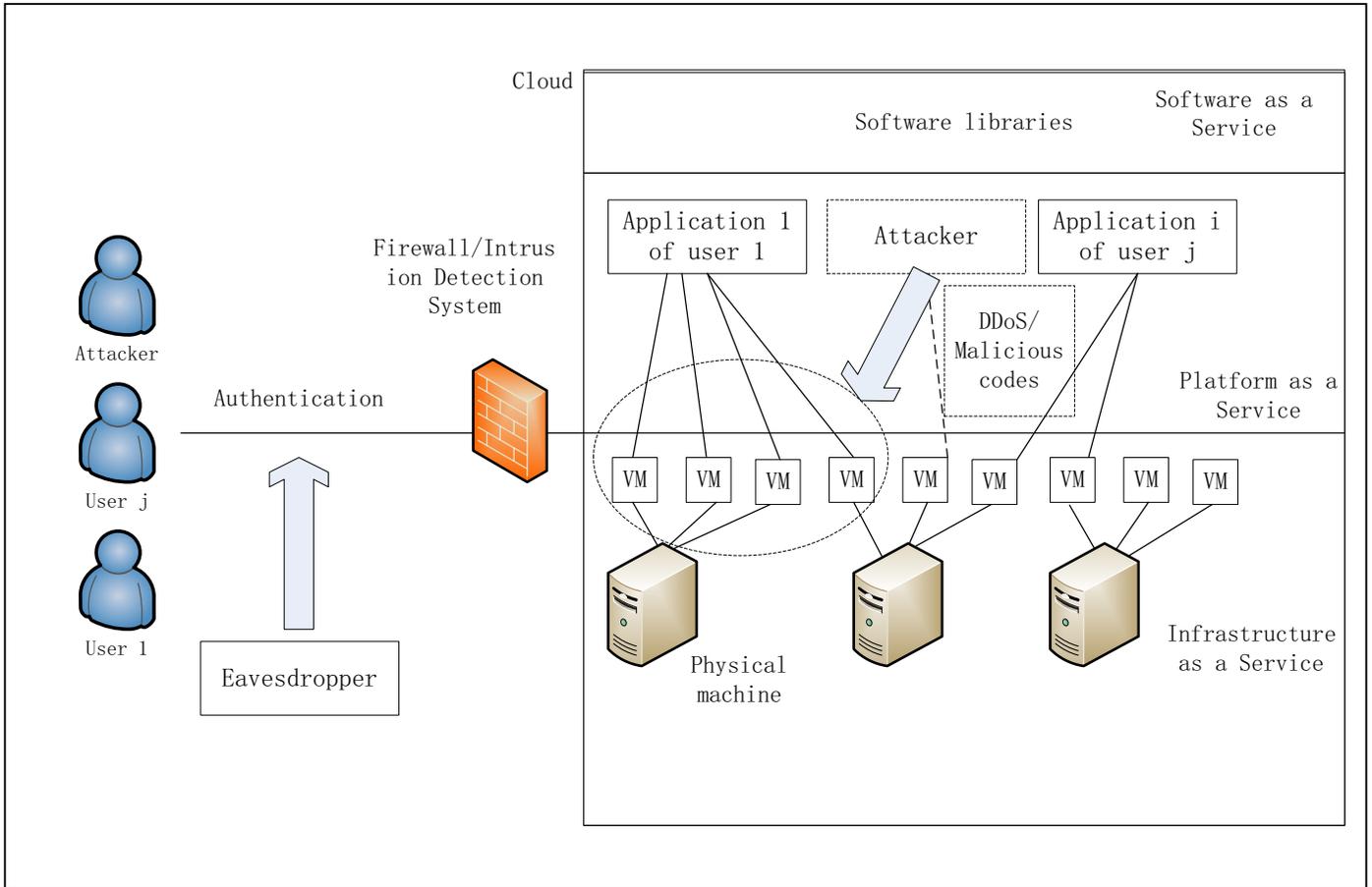

Fig. 1. Cloud Infrastructure and Possible attacks in different points

Table 1. summarizes the possible attacks and solutions. The next few paragraphs discuss about possible attacks at different points and protection methods in detail.

TABLE 1  ATTACKS IN CLOUD

| Attacks | Target and effect | Protection methods |
|---|---|---|
| Eavesdropping | Users. key pair for authentication would be obtained by adversaries in the middle | Regular key pair updates, multifactor authentication |
| Malicious Codes | Users and Cloud Provider. SaaS cloud malicious codes in application would propagate | Distributed to multiple machine instances in the same cloud |
| Virtual Machine | Users. virtual machine storing data would be compromised | Virtual machine segregation |
| DDoS | Users and Cloud providers, virtual machine instances containing | Virtual machine migration, Virtual Private Cloud |
| Insider Attacks | Users, data in cloud would be leaked to adversaries | Distributed storage, Encryption |
| Data Integrity Attacks | Users, data in cloud would be leaked to adversaries | General Data Protection Regulation (GDPR), Intel SGX |

A. *Authentication*

Attacks during authentication occur between end users and cloud environment. Most public cloud provid-ers today like AWS use public/private key for authentication. Users would first login to their account with username and password, then create and download key pair generated. For further connection and authentication to EC2 instances, the key pair is used. However network eavesdroppers in the middle would intercept the key pair and do cryptoanalysis or man in the middle attack. Updating key pair used regularly and using multifactor authentication might be helpful in this case.

### B. Virtual Machine Attacks.

In public cloud, multiple tenants would share the same physical machine through virtualization. If adversaries pretending as legal users are able to login to the virtual machines, they would be able perform attacks for the following reasons:

a) Calls to a virtualized network device would be passed to the physical network device. If malicious code is inserted, it would propagate and affect other virtual machines on the same physical device or even other physical machines.

b). Adversary might exploit relaxed access control and inter VM communication on the same physical machine to perform attacks

### C. DDoS attacks.

In perspective of cloud providers, it would be hard for attackers to perform DDoS attacks due to huge amounts of servers. However, in perspective of users, adversary would be able to make the specific servers which contain that user's data unavailable to use if they know the location of data residence. One could solve this by live virtual machine migration. Also cloud providers like AWS enables Virtual Private Cloud (VPC) service for stronger access control

### D. Insider abuse.

Unlike in single machine and clusters, machines in cloud computing are in possession of cloud providers. Cloud providers would clearly know the data contents, location, and computation/analytics methods performed over data. If cloud providers collude with adversaries, data stored would be exposed. Thus methods to protect data security in untrusted cloud providers need to be developed. Several existing methods like multiple storage are discussed in the following sections.

## III. CLOUD DATA SECURITY CHALLENGES

As mentioned above, there would be more concerns on data security in cloud environment than in traditional single ma-chine which are in hold of users themselves. In this section, different concerns and possible attacks during cloud data usage and storage phases are summarized.

### A. Data Use

Once data is migrated to cloud, cloud providers will clearly get hold of everything users transferred to cloud machine in-stances. Both adversaries and cloud providers themselves might misuse the data stored in cloud. Thus some kinds of data transformation might be needed to prevent valuable information leakage.

If cloud is only used for data storage and no further operations are needed, simple encryption is feasible. However, in most cases, further processing might be needed.

Users might need to apply certain processing over the data stored. For example, computation might be needed like matrix multiplication. Also data analytics methods such as ma-chine learning algorithms need to be applied for data classification.

Data computation and analytics is within the cloud itself. However certain use of cloud data will also need the interaction between local users and remote cloud. For example, users might need to retrieve certain data for modification through data querying.

Concern about the above cloud data usage is described in details as follows:

1). Processing on encrypted data. For data and computation intensive applications, the algorithms themselves also reside in cloud. Cloud providers could infer from the algorithms what kinds of data are stored. Besides, to protect data, the data to be processed also need to be encrypted. How to process encrypted data without decryption remains a question.

2). Query analytics attack. In cloud environment, large volumes of data are stored. Re-mote users need to query data hosted in cloud for auditing, processing or other operations. Queries could be utilized to perform attacks for the following reasons:

a). These queries and queried results would go through Internet connection where eavesdroppers would sit in the middle to obtain these results.

b). Cloud provider themselves are untrusted. The query processing procedures are transparent to cloud providers. Thus even if the original data is transformed and stored inside, and the same query is transformed to different values each time and sent to the cloud environment, cloud providers would still obtain some information from it.[18]

c). Priori knowledge about the data which could be obtained from open resources could be combined with the analytical results to infer valuable information.

### B. Data Storage

In cloud environment, users' data are stored in remote virtu-al machine instances in possession of cloud providers. Ac-cording to [1], there could be various outside attacks over virtual machines including malicious codes attack, compromising the corresponding Virtual Machine Monitor etc. Besides the outside attack, users lack of physical control of their data. Insiders of cloud providers could clearly see what is stored in their virtual machine instances. It would be a catastrophe if the insiders of cloud providers collude with adversaries to intentionally modify or leak customers' data.

Data storage security includes confidentiality, integrity and availability. For data confidentiality, how to prevent information leakage and efficiently check data integrity over large amount of data stored in cloud remains a question. The goal here is to minimize the probability to recover the origi-nal data obtained from the compromised cloud storage sys-tem.

For data integrity, adversaries as well as cloud providers would modify the data intentionally. An efficient integiry checking over large amount of data is necessary.

As of data availability, both system maintainance and at-tacks would cause customers' data to be unavailable.

.

## IV. DATA CONFIDENTIALITY AND AVAILABILITY PROTECTION METHODS

### A. File distribution in multiple storages.

Multiple storages are applied to minimize the information leakage when a single storage is compromised. With this method, encryption is not needed. Table 2 summarizes several existing methods of multiple storages.

TABLE 2 MULTIPLE STORAGE METHODS

| Splitting Methods | Distribution | Reconstruction |
|---|---|---|
| Maximum relative entropy splitting[16] | To multiple machine instances in the same cloud | All the pieces of data |
| Polynomial interpolation[19] | To multiple clouds | Only k out of n pieces of data |
| Divide and Conquer[21] | Random distributed to multiple cloud providers | All the pieces of data |

[16] proposes an optimal data splitting and distribution algorithms to minimize the useful informational content contained in each file chunk stored in different virtual storage. For each file chunk $c_i$ and the entire file f, the goal of splitting is to have the largest relative entropy $I(f,c_i)$, which is the information lost when $c_i$ is used to approximate f. In order to reconstruct f, one needs to find the correct set of virtual storages among all the virtual storage volumes, and the correct sequence of file chunks among all the file chunks in that set of virtual storage volumes. For insider attack, one would know the storage set, but not the sequence. The distribution of files is to minimize the successful recovery probability.

However, there are certain problems for this method: a). Additional information like the index of each chunk needs to be stored in private cloud. b). The corruption of single piece of data would lead to the corruption of the entire file, additional backups are needed. Tradeoff between confidentiality and availability exists. c). The splitting algorithm proposed is not efficient, instead dynamic programming could be used. d) Data processing is not discussed in this paper.

In [19], the secret sharing scheme "(k, L, n)-threshold scheme" is used to compute and distribute the file to different clouds, where k is the required number of shares needed to recover the original file, L is the data size of saved file in each cloud, and n is the shares of files to be distributed. The original data D is divided into $D_i$ through a random k-1 degree [20].

$$q(x) = a_0 + a_1 x + a_{k-1}x^{k-1} \qquad (1)$$

where $a_0$=D, $D_n$=q(n). Compared with method in [16], no additional information is needed to store in local machine. File could still be recovered with at most n-k pieces of file corruption. However, there are still problems with this method: a). Rely on a single cloud is not a good idea, since cloud provider is untrusted, and they could easily recover the file by getting all the pieces of file chunks stored in it. b). It has high complexity if the file size is in order of Terabytes.

In [21], the hierarchical organization from bottom to top is data storage nodes, data processing nodes and a Command and Control node which is needed to keep track of which piece of file resides in which storage node. A small subset of data which contains little information resides in each leaf storage node. There are still several problems with this method.

a) Data processing is mentioned in this method, however it is before storage, and also it needs decryption first before processing.

b) The splitting method simply divide file to extremely small pieces. This minimize the informal contents contained in each piece but needs sufficiently large number of machine instances compared to method [16].

### B. Processing over encrypted data

In the above section, multiple storages are applied to secure data confidentiality by minimizing information contained in each storage node. However, in the case of data computation, the above method is not applicable. Since for data processing in cloud, each piece of data needs to contain computable information. A better solution might be computing while keeping data encrypted.

Craig Gentry [22] first proposes a "fully homomorphic encryption" (FHE) scheme to compute over encrypted data. FHE is a scheme that operates on ciphertexts so as to add, subtract, and multiply the underlying messages. It consists of key generation, encryption, decryption and evaluation algorithms. Each evaluation function is associated with a function $f(m_1,m_2,…,m_t)$ which could be represented as a computable combination of original texts $m_1,m_2,…,m_t$. This encrypted function f is what is sent to cloud for computation. This method has the following problems: 1) FHE needs running evaluation algorithm on the decryption function of a constructed bootstrappable homomorphic encryption scheme first. However, this is computationally expensive. 2) It is less efficient than lattic-based scheme.

Later on, this FHE method is improved. [24] brought up a probabilistic decryption algorithm that can be implemented with an algebraic circuit of low multiplicative degree to enable faster FHE. In [25], FHE protocol for multiple users is designed. In [27], detailed implementation of basic operators including addition, subtraction, multiplication, division, relation is discussed with logical circuits and algorithms. Based on that, data structure like array, link list, stack, queue etc., and the corresponding operations to these data structures are discussed. The timing evaluation is given in the end.

Built on top of homomorphic encryption, more advanced processing techniques are developed. In [23], CtyptDB is proposed which enables SQL query processing over encrypted data. It is under the following assumptions 1). Query and query results would not be changed. 2) The proxy issuing queries including the encrypted data is reliable. 3) The DBMS is not trusted. This applies to cloud environment since the DBMS in possession of cloud providers could not be trusted. The difference is that queries would be intercepted by adversaries sitting in the middle between users and cloud environment.

In CryptDB, adjustable query based encryption is applied. That is, different operation sets have different encryption security strength layers. And in the original table, each column would be transformed with different encryption methods involving Random, Deterministic, Order Preserved, Homomorphic, Join and Search encryption in different operation sets. Operation sets include Equation, Order, Search and Addition sets. One method of Random encryption is by using chaining cypher with a different random initialization vector each time. In [18], a detailed Order preserved encryption with random space perturbation is introduced. This is to expand the record with two additional dimensions by a deterministic dimension and random generated dimension.

Similar to CryptDB, in [27], MONOMI is designed to query over encrypted data. However, compared to CryptDB, it could afford more analytical queries.

Moreover, Raphael Bost et.al [28], construct hyperlanes decision, Naive Bayes, and decision tree classification over encrypted data based on additively homomorphic encryption with public/private key scheme. MrCrypt[31] provides static analysis for secure cloud computations. Crypsis[32] has practical confidentiality preserving Big Data Analysis

## V. DATA INTEGRITY PROTECTION METHODS

According to [30], data integrity attacks in cloud includes Data Modification Attack, Tagforgery and Data Leakage Attack, Replay and Timeliness Attack, Roll-Back Attack and Collusion Attack and Byzantine Attack.

Cong Wang et al [26] achieve data integrity by utilizing precomputed token. The original file is represented by m column vectors in Galois Field. This file would be encrypted by multiplying with a certain matrix to achieve additional k parity check vectors. Thus, the encrypted file contains m+k columns in total. Further computation would be done over the last k parity check columns to protect the confidentiality. At the end the encrypted m columns and the modified last k parity check columns are sent to store in cloud. The encryption has additive homomorphic encryption attributes, which enables efficient updates of the file.
On each encrypted vector, if t times of verification is need-ed. Each time, a token is computed using partial blocks of data in the vector, thus (m+k)t tokens are precomputed in total. In order to verify the integrity of the file, the index would be sent to cloud storage, and same computation procedure on that partial data is done to generate the signature, which would be sent back and compare with the original token. Thus, if file corruption occurs, the corrupted location would be known.
This is mathematical approach of integrity checking is in data level. However, there are certain problems with this method:

a). All precomputed tokens need to be stored in local environment. Although the paper mentions that it could be stored in remote cloud, the untrusted cloud providers would be able to modify it.
b) Since the tokens are not generated on all parts of the file, it could only provide probabilistic integrity assurance.

In [14], a Trusted Cloud Computing Platform (TCCP) based on trusted computing is proposed to protect confidentiality and integrity. The trusted platform module is implemented in each node. However, since users don't have control over the physical machines, remote attestation is needed to ensure that measurement indeed comes from the VM which users are running applications on. In each virtual machine, trusted platform module (TPM) is embedded and a trusted virtual machine monitor (TVVM) is installed during the booting. Besides, an external trusted coordinator (TC) is used to do the attestation. The virtual nodes need to register with TC.
According to [29], Intel already has Intel Trusted Execution Technology (TXT) based on the TPM. It is compatible with OpenStack which is a open-source software platform for cloud computing. In it, and OpenAttestaion server is responsible for communicating with the trusted computing pool of hardware and software.
However, with trusted cloud computing, application level attack would not be determined. For example, if data stored in database are compromised. It would not be detected. In order to use TCCP, additional application level security needs to be implemented

## VI. DATA PRIVACY PROTECTION METHODS

In [4], data privacy against data mining is kept by distributing data to different cloud providers. Thus, data analytics based on each part in one cloud might be misleading. For example, prediction made on the overall data file could be different from that made on each part. However, this kind of approach would not protect each individual's sensitive data. For example, a database contains columns of username and the corresponding income. If the file is simply divided by rows as in [4], each individual's income information is still leaked.

In [9], anoymization is applied. A unique indexing of each row in databased is generated by hashing of the unique identifiers of each row. The hashing information needs to be kept locally, and table after removal of these unique identifiers would be split by columns to different cloud providers. In this way, individual privacy is kept, however since a whole column will be stored in one cloud, possible data mining attacks could be applied to predict useful information.

In [18], a scalable local recoding method is proposed, which could minimize data distortion. However, by generalizing the records, queries over stored data might not be that effective.

## VII. DATA SECURITY AND PRIVACY ARCHITECURE WITH PROTETION METHODS

In this section, a cloud computing architecture involving multiple clouds with data security and privacy protection is proposed..

### A. Threat model and System Architecture

As shown in Figure 2, multiple cloud systems are applied. The data is divided to be sent based on its secret level and operations needed over the data. The secret level includes top secret, secret and unclassified. The operations on data includes no operations, basic operations including simple addition, subtraction, multiplication and division, and advanced analytics

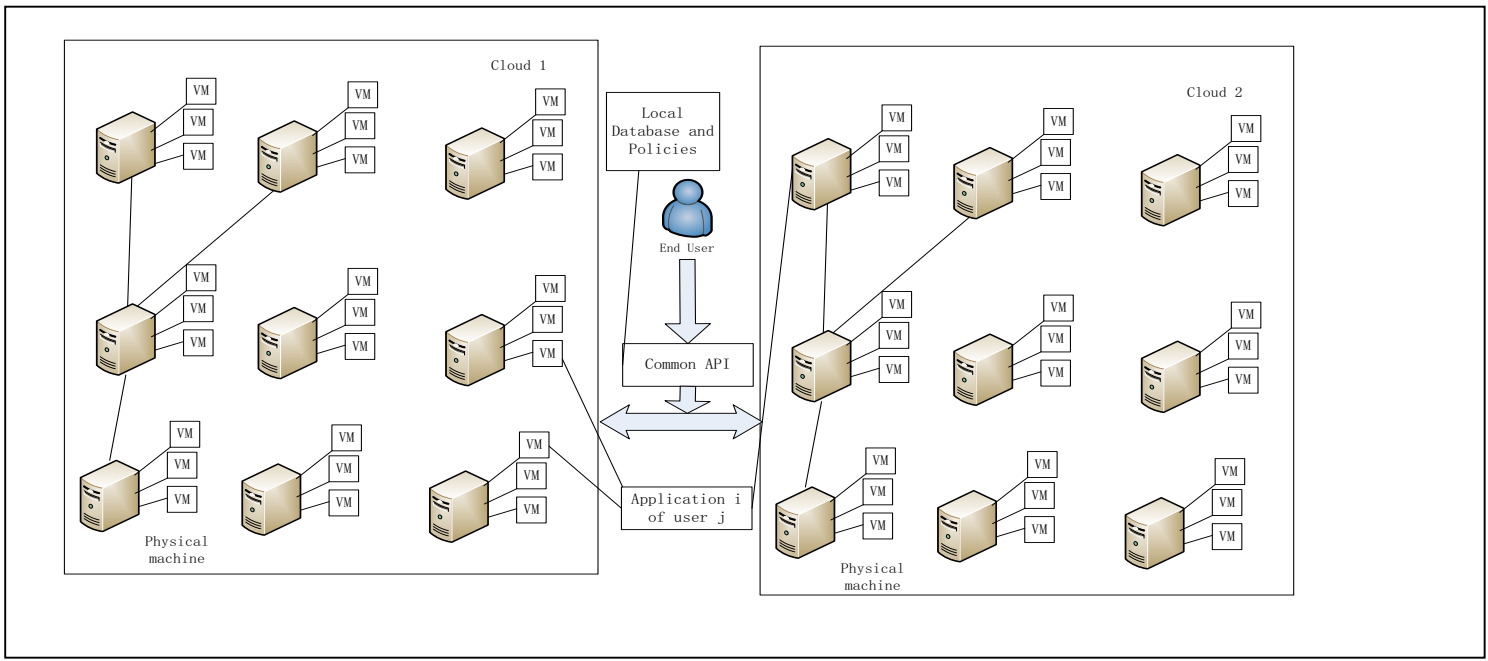

Fig. 2. Multiple Hierarchical Cloud infrastructure with a common API

Within each cloud, a hierarchical organization of machine instances is applied. According to [1], based on Bayesian probabilistic access from bottom to top layer machines, data in the highest level has the highest confidentiality.

Here, a PaaS cloud model is considered. As shown in the figure, data could be sent to multiple layers of machine instances across different cloud providers. The cloud environment itself would apply trusted cloud computing discussed above.

It's known that there is not a system which could be fully trustworthy. Tradeoff exists among cost, efficiency and security. For top secret level, it is better to store locally, and for unclassified data, it could be stored directly to a single cloud platform without encryption. In Table 3, different protection mechanisms over secrete data are summarized.

TABLE 3
PROTECTION OF SECRET DATA WITH DIFFERENT OPERATIONS

| Operations | Representative applications | Possible Protection methods |
|---|---|---|
| No operations | Backup service, e.g. dropbox | Multiple storage [16] |

| Basic Operations | Database | CryptDB [17] |
| --- | --- | --- |
| Advanced Analytics | Machine learning | Additively homomorphic encryption[28], Cripsis [32] |

In this architecture, the local user side is considered as trusted party, the interaction between user and cloud, and the multiple cloud providers themselves are all untrusted.

In terms of different secret level and operations needed, a common preprocessing API is defined for anonymization, authentication, processing data with different operations and secret level etc. The input to the common API is data to be sent, secret level and computation difficulty. Upon receiving a request from user, the common API would further call the functions in libraries to perform the correspond operations. It would refer to and update local database. For example, if an input is (data, secret, no operations). The common API would further call the data splitting and distribution library and also update the local database to store the file chunk location information.

### B. Security Evaluation

For different cloud providers, its ranking and cost are different. Also for different methods, the overall performance including time complexity T, cost C, security S and privacy P level are different. In order to evaluate a method with the overall performance, a weighted linear ranking is proposed as follows:

$$Pr = a_1 \times T + a_2 \times C + a_3 \times S + a_4 \times P \quad (2)$$

where $a_i$ is adjustable weight. The time complexity needs to be evaluated per algorithm base. The cost could be measured through the billing system of cloud providers. For data security and privacy in cloud, it is kind of overlapping. Because if encryption is used, as long as the data is confidential, the sensitive information will not be disclosed. In addition, the data mining attacks would be mitigated. Thus, here only data security is considered.

Data security includes data availability, integrity and confidentiality. Data availability would be compromised in two main categories in cloud. First is the cloud architecture reliability, i.e. the regular maintenance and failure of machine instances. The other comes from attacks talked above.

For data confidentiality, there are three layers of confidentiality in the proposed architecture. In order to recover data, one needs to bypass the authentication of a cloud platform. Inside the cloud platform, hierarchical access to data is required. The encrypted information disclosed is partial. Thus, the non-confidentiality Level Probability of Accessing one of the cloud service * Hierarchically Access Probability that the original information contained in the virtual machine.

## VIII. CONCLUSIONS AND FUTURE WORK

### A. Conclusions

This paper focuses on the data storage and computation security and privacy. Different methods are compared, problems and advantages with the existing methods are discussed. A hierarchical multi-cloud architecture with a common preprocessing API and local database to deal with anonymization, authentication, processing data with different operations and secret level is proposed.

### B. Future work

The proposed architecture could provide various security methods according to the data type and usage to reduce the complexity by calling a common API. However, there are some problems with this design to be solved:

a). The call to the common API would be intensive, thus caching and scalable procedures are needed.

b). Although most storage and computation are done in remote cloud, the preprocessing and certain data related information need to be stored and maintained locally.

c). Since algorithms over encrypted data are applied, customer-oriented algorithms need to be developed. Users need have a good knowledge of encrypted data.

d). A fully functional benchmark for the system need to be designed, implemented and evaluated with quantitative and qualitative performance metrics.